\documentclass[sigconf]{acmart}

\usepackage{booktabs,multirow} 
\usepackage{graphicx}
\usepackage{caption}
\usepackage{subcaption}

\graphicspath{ {./figures/} }

\copyrightyear{2019} 
\acmYear{2019} 
\setcopyright{acmlicensed}
\acmConference[MM '19]{Proceedings of the 27th ACM International Conference on Multimedia}{October 21--25, 2019}{Nice, France}
\acmBooktitle{Proceedings of the 27th ACM International Conference on Multimedia (MM '19), October 21--25, 2019, Nice, France}
\acmPrice{15.00}
\acmDOI{10.1145/3343031.3351050}
\acmISBN{978-1-4503-6889-6/19/10}

\author{Hsin-Ying Hsieh}
\affiliation{%
  \institution{National Chiao Tung University}
  \city{Hsinchu}
  \country{Taiwan}
}
\email{x4ga71@gmail.com}

\author{Chieh-Yu Chen}
\affiliation{%
  \institution{NVIDIA Corporation}
      \city{Taipei}
  \country{Taiwan}
}
\email{jaych@nvidia.com}

\author{Yu-Shuen Wang}
\affiliation{%
  \institution{National Chiao Tung University}
  \city{Hsinchu}
  \country{Taiwan}
}
\email{yushuen@cs.nctu.edu.tw}

\author{Jung-Hong Chuang}
\affiliation{%
  \institution{National Chiao Tung University}
  \city{Hsinchu}
  \country{Taiwan}
}
\email{jhchuang@cs.nctu.edu.tw}

\begin{document}
\fancyhead{}

\title{BasketballGAN: Generating Basketball Play Simulation \\ Through Sketching}

\begin{abstract}
We present a data-driven basketball set play simulation. Given an offensive set play sketch, our method simulates potential scenarios that may occur in the game. The simulation provides coaches and players with insights on how a given set play can be executed. To achieve the goal, we train a conditional adversarial network on NBA movement data to imitate the behaviors of how players move around the court through two major components: a generator that learns to generate natural player movements based on a latent noise and a user sketched set play; and a discriminator that is used to evaluate the realism of the basketball play. To improve the quality of simulation, we minimize 1.) a dribbler loss to prevent the ball from drifting away from the dribbler; 2.) a defender loss to prevent the dribbler from not being defended; 3.) a ball passing loss to ensure the straightness of passing trajectories; and 4) an acceleration loss to minimize unnecessary players' movements. To evaluate our system, we objectively compared real and simulated basketball set plays. Besides, a subjective test was conducted to judge whether a set play was real or generated by our network. On average, the mean correct rates to the binary tests were 56.17 \%. Experiment results and the evaluations demonstrated the effectiveness of our system. Code is available at \url{https://github.com/chychen/BasketballGAN}.

\end{abstract}



\maketitle
\section{Introduction}
\label{sec:intro}

Basketball is a popular sport that is played or watched by hundreds of millions of people in the world. It is a five against five players game, in which the goal is to score points by throwing the ball into the basket hoop. To prevent defensive stops by defenders, adopting tactics that can create an open space for the shooter is important. In practice, basketball coaches draw up players' movements and ball passing routes on a tactic board to show players how to execute an offensive tactic. After reading the sketched tactic, professional players would simulate the play based on their own intuition and experience, and then execute the tactic on the court as the coach drew up. However, this is not always the case for novice players. The play could differ a lot to the tactic expected by a coach due to the influence of defensive players on the court.

\begin{figure*}[t]
    \centering
    \includegraphics[width=\linewidth,keepaspectratio]{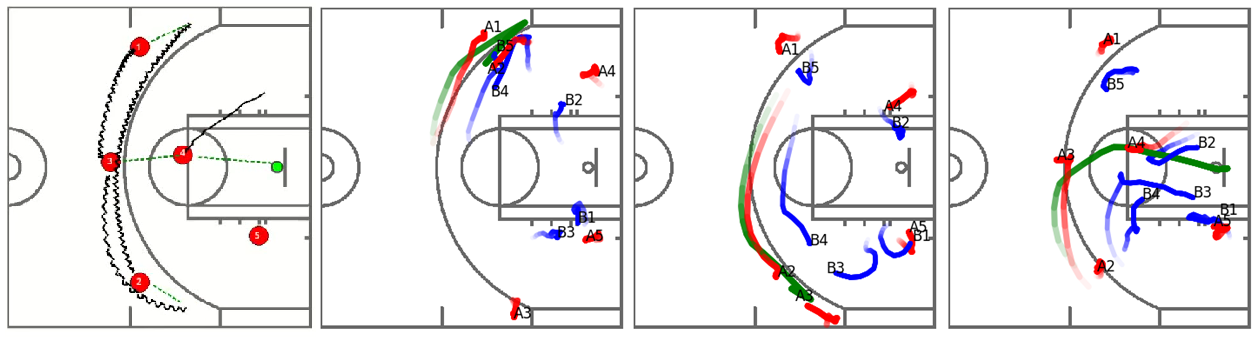}
    \caption{(Left) A sketched basketball offensive tactic. Green and red circles are the ball and the offensive players, respectively; zigzag, smooth, and dotted lines indicate the movements of the dribbler, non-dribblers, and the ball pass, respectively. (Middle left to right) The simulated basketball set play based on the sketched tactic. Offense, defense, and ball are coded in red, blue, and green, respectively. We segment the set play into three consecutive segments for clear visualization. Besides, trajectories from transparent to opaque indicate the movement direction. We recommend readers to watch our accompanying video for the sketching and the simulation results because the dynamic animations are difficult to visualize in still images.}
    \label{fig:Coach_sketchComp}
\end{figure*}

The main difficulty of executing an offensive tactic is not knowing how the opposing team will defend the tactic. Figure \ref{fig:Coach_sketchComp} is a simple ball rotation tactic designed by a coach, in which the goal is to confuse the defenders by moving the ball from top to bottom in a quick fashion. Because the defensive players' movements are unknown, it may not be easy to understand how and why the tactic is effective. Novice players would be less confident and dubious to the tactic when playing basketball games on the court. As a result, the flow of their offence has a higher probability to be disrupted by the opposing team. Even when the offense is performed by professional players, it is not guarantee that the tactic will succeed because the opposing team may react by adopting the right defensive tactic. Therefore, a system that can simulate and predict how the opposing team will react would be very helpful to players and coaches. Specifically, novice players can better understand the tactic and be more confident when they execute the tactic on the court. In addition, professional players and coaches can evaluate and determine whether an offensive tactic is effective when it is used against the opposing team. The simulation based on historical data can back up their intuition and help them make right decisions.


Simulating a defensive play based on the offensive condition is not an innovative idea. Chen \textit{et al.} \cite{chen2018generating} presented a generated adversarial network (GAN), which takes the ball and the offensive players' movements as input, can achieve the goal. However, the system is infeasible because real movements of the ball and the players are difficult to sketch on a tactical board. Seidl \textit{et al.} \cite{seidlbhostgusters} transformed an offensive tactic sketch into an animation sequence and fed the sequence to a long short term memory network to simulate the defensive play. Since sketch lines are often smooth and lack temporal dynamics, the simulated plays are unrealistic -- players move with an identical speed on the court. In addition, wide opens frequently occur in their results.

We present a GAN that is composed of convolution layers to simulate basketball set plays. The input of the network is an offensive tactic sketched by coaches and the output is a basketball set play, in which the offensive part fulfills the input condition. The presented GAN is composed of a generator and a discriminator. The former takes the condition and a random noise as input to generate basketball set plays, and the latter evaluates the realism of the plays. We train the network on a basketball player movement dataset released by NBA. The goal is to minimize the adversarial loss measured by the discriminator. In addition, to improve the realism of the simulation, we applied the dribbler loss, defender loss, ball passing loss, and the acceleration loss to guide the network training. These heuristic loss functions effectively prevent abnormal player behaviors on the court. Considering that users often draw smooth lines to specify an offensive tactic, and the lines lack temporal dynamics, we prepare the offensive conditions by spatially and temporally smoothing real players' movement trajectories. The network is trained to construct details between sketches and real offensive players' movement trajectories, and the missing defensive set play. After training, users can sketch an offensive tactic and forecast how the opposing team will defend the tactic. 

We trained the network to simulate basketball set plays accordingly to the sketched offensive conditions. To evaluate the results, we compare real and simulated basketball set plays in terms of the ball and the players' movements, and their relationships. The results show that the basketball set plays simulated by our system are realistic. In addition, since quantitative evaluations may not be perfect, we also conducted subjective tests with 50 participants (6 professional players, 22 NBA fans, and 22 ordinary people) to evaluate the results. Participants were asked to identify the real play from two samples. On average, their mean correct rate to the binary tests were 56.17 \%.  


\section{Related work}
\textbf{Basketball analytic} has been a research trend for years. Early works attempt the use of statistic and number crunching in determining which player statistics are crucial and influence the outcome of a basketball game \cite{oliver2004basketball, kubatko2007, sampaio2006, sampaio2010}. Franks \textit{et al.} presented a system \cite{franks2015counterpoints} to determine which player is a better defender on specific locations on the court, and how efficient the player is at defending an offensive player. Beshai \cite{beshai2014buckets} implemented a system to visualize basketball shot data in the NBA. It provides users a platform to enjoy basketball statistics through interactive charts and player comparisons. 

In the past few years, the existence of STATS SportVU player tracking data has opened new doors for researchers to examine players' movements. Applications, such as tactic classification \cite{chen2015, wang2016} and shot make prediction \cite{harmon2016predicting, cervone2016multiresolution, shah2016} in basketball set plays, were presented by analyzing players' movement trajectories. In addition, several works were introduced to generate basketball set plays. Toronto Raptors showcased a private system developed by their analytic team, which could use players attribute and skill-set to determine where each individual defensive player should stand on the court while on the defensive end. Seidl \textit{et al.}\cite{seidlbhostgusters} presented an interactive sketch play system. When given an input of offensive trajectories, their model can synthesize ghosting players that imitate defensive behaviors in basketball games. Chen \textit{et al.} \cite{chen2018generating} trained a GAN that takes a real offensive play as input and generates the corresponding defensive play for coaches and players to consider. Although the works of Seidl \textit{et al.}\cite{seidlbhostgusters}, Chen \textit{et al.} \cite{chen2018generating} and ours attempt to achieve a similar goal, their systems focus on only defensive plays. They either assumes that offensive players move smoothly on the court or that realistic offensive set plays can be easily specified. In contrast, our system simulates realistic and complete basketball set plays according to simple sketches.

\textbf{Generative adversarial network} (GAN) is an unsupervised deep generative model pitting a generator and a discriminator against each other to generate an output $G(z)$ from a given latent noise $z$ \cite{goodfellow2014generative}. A GAN learns to generate realistic samples from a dataset distribution, as it has been used widely in content generation ranging from images, videos \cite{li2018video} to music \cite{mogren2016c} and text \cite{fedus2018maskgan}. Because the input $z$ is uncontrollable, a conditional version of GAN was then introduced \cite{mirza2014conditional}. By feeding a $y$ condition into the network, we can specify a specific output rather than a generic sample generated from a random noise. Accordingly, many works have taken advantage of conditional GANs, such as generating images based on text description \cite{reed2016generating}, style transfer of real image into a desired style \cite{chen2018cartoongan} or image to image translation \cite{isola2017image,zhu2017unpaired}.  

Although GANs are promising and can achieve exceptional results, the network itself is difficult to train and experimentally suffers a few flaws such as models never converging or mode collapse \cite{arjovsky2017towards}. To improve learning stability, Ajovsky \textit{et al.} \cite{arjovsky2017wasserstein} presented Wassestein GAN (WGAN), which applies the Earth Moving distance to measure the similarity of real and fake samples. Later on, Gulragini \textit{et al.} \cite{gulrajani2017improved} improved upon the works of WGAN by gradient penalty and demonstrated the network is over its predecessor on quality and convergence. In addition to WGAN, there have been also several strategies presented to improve network training, such as energy-based GANs \cite{zhao2016energy}, minibatch discrimination \cite{salimans2016improved}, boundary equilibrium GANs \cite{berthelot2017began}, and spectral normalization \cite{Miyato+:2018:SpectralNF}. 



\section{Basketball Games Simulation by Sketching}
We train a GAN to simulate basketball set plays according to the sketched offensive conditions. The network learns to construct details between sketches and real offensive players' movement trajectories, and the missing defensive set play. Details are described in this Section.



\subsection{Training Data}
The work dataset comes from Player Tracking data provided to the public from STATS SportVU. It contains NBA games during half of the regular 2015-2016 season (around 600 games played). The ball and the ten players' positions on the court were captured at 25 frame per second. We down-sample the tracking data to 5 frames per second for training. In addition, we segment a whole game into offensive plays that start when the ball is dribbled across or in-bounded from the half court, and ends when a shot is made or missed. To achieve said result, the NBA play-by-play information \footnote{NBA play-by-play text information records every event type that occurs during a game by officials. Every event is a basketball play that occurs on the court, e.g. rebound, assist, shot made, shot attempted.} is parsed to extract every shot made or missed, which can be done by matching the time of the play-by-play and tracking data since both datasets provide a time variable.

We prepare offensive conditions by imitating user sketches of those NBA player tracking dataset. The sketches are often smooth lines from a starting point to a desired end point, unlike real player movement trajectories, where players may move in unpredictable ways. In addition, the sketched movements in each segment are of uniform speeds because of the lack of temporal information. To synthesize sketched set plays, each real play is first segmented into segments by ball pass or ball shot at the basket, as these actions often indicate the end of a phase during a sketch play design. Then, each player movement trajectory is represented by a B\'ezier curve, in which the control points are simplified from the original tracking data by the Ramer-Douglas-Peucker (RDP) algorithm. The segment length is retained in this new representation. Different to players' movement trajectories, the ball trajectory is identical to the handler's trajectory when dribbled, and only moves in a straight line to an intended receiving player when passed, or to the basket hoop when shot.


\subsection{Sketching Interface}
We provide users with a graphical interface to sketch and design offensive set plays. The interface comes with a half-court basketball board, with five movable circles that represent the offensive players. Figure \ref{fig:Coach_sketchComp} (a) left shows an example. To sketch a play, users first set up the five players' positions, and then double click a circle to indicate the dribbler. By holding on a circle and moving it, the player's movement trajectory is sketched upon the board, displaying how the player moves from start to finish. Users can specify a ball pass by clicking the dribbler and then the intended receiver on the court. They also can create a shot in a similar fashion. Instead of an intended receiver, the sketch is directed towards the basket hoop.

The sketched set play lacks precise temporal information. Namely, the start and the end frame of a trajectory is unknown; and the player's movement speed is uniform. We consider the interfaces used in popular basketball apps and the work presented by Seidl \textit{et al.} \cite{seidlbhostgusters}, and then decide to temporally segment an offensive play by ball pass and ball shot events. Each segment length (i.e., frame number) is determined by the longest trajectory and the mean player's movement speed, which is analyzed from the NBA player tracking data. In addition, we smooth the sketched trajectories in a similar fashion as the training data, and then determine the player position in each frame by uniform sampling. This approach can reduce the gap between real and synthesized sketches.


\subsection{Generative Adversarial Network}
We train a GAN to simulate basketball plays that can fulfill the offensive conditions sketched by users. Specifically, the input to the network is a set of sketch lines that condition the movements of a ball and five offensive players. The output is a complete basketball play that contains the movement trajectories of a ball, five offensive players, and additionally five defensive players. We recommend readers to watch the accompanying video for better understanding the input and output of the network because basketball plays are dynamic and difficult to visualize in still images.

\subsection{Network Architecture}
\label{sec:GAN_arch}
\begin{figure}[t]
    \centering
    \includegraphics[width=\linewidth,keepaspectratio]{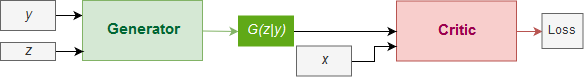}
    \caption{Network Overview. Offense condition $y$ and latent noise vector $z$ are used to simulate a basketball set play $G(z|y)$. The simulated play is then evaluated by a critic network. The generator and the critic are trained to compete against each others.}
    \label{fig:network_overview}
\end{figure}

\textbf{Architecture overview.}
The presented GAN consists of a generator and a discriminator competing against each other, as illustrated in Figure \ref{fig:network_overview}. The generator $G$ takes an offensive condition $y$ and a latent noise vector $z$ as input and generates the corresponding basketball play $G(z|y)$. Specifically, $y$ is a $t \times 18$ matrix, where $t$ is the frame number of a basketball play and each frame is conditioned by an $18$ dimensional feature vector. Twelve dimensions of this vector are the ball and the offensive players' positions $\mathcal{R}^2$ on the court. Six dimensions of the vector provide the ball status. When the ball is in possession of player or when a shot attempt occurs, a $1$ is encoded to the corresponding player or the basket hoop. Otherwise, a $0$ is encoded. Accordingly, at most one dimension of the ball feature can be $1$; and a zero vector indicates a ball pass. The output of the generator $G(z|y)$ is a $t \times 28$ matrix, where $t$ is the frame number the same to the input and each frame contains the positions of a ball and ten players, and a ball feature. By requesting the generator to generate its own ball feature, we can impact the generator to keep the ball near the handler when the ball is not passed or shot.

The discriminator evaluates the realism of a basketball set play $G(z|y)$. Specifically, it criticizes if the offense fulfills the given condition $y$, if both offence and defence together are realistic, and if the generated ball feature displays the correct ball status. Therefore, we form the real and fake pairs by $y \oplus x$ and $y \oplus G(z|y)$, respectively, to train the discriminator, where $\oplus$ indicates concatenation.

\begin{figure}[t]
\centering
  \includegraphics[width=\linewidth]{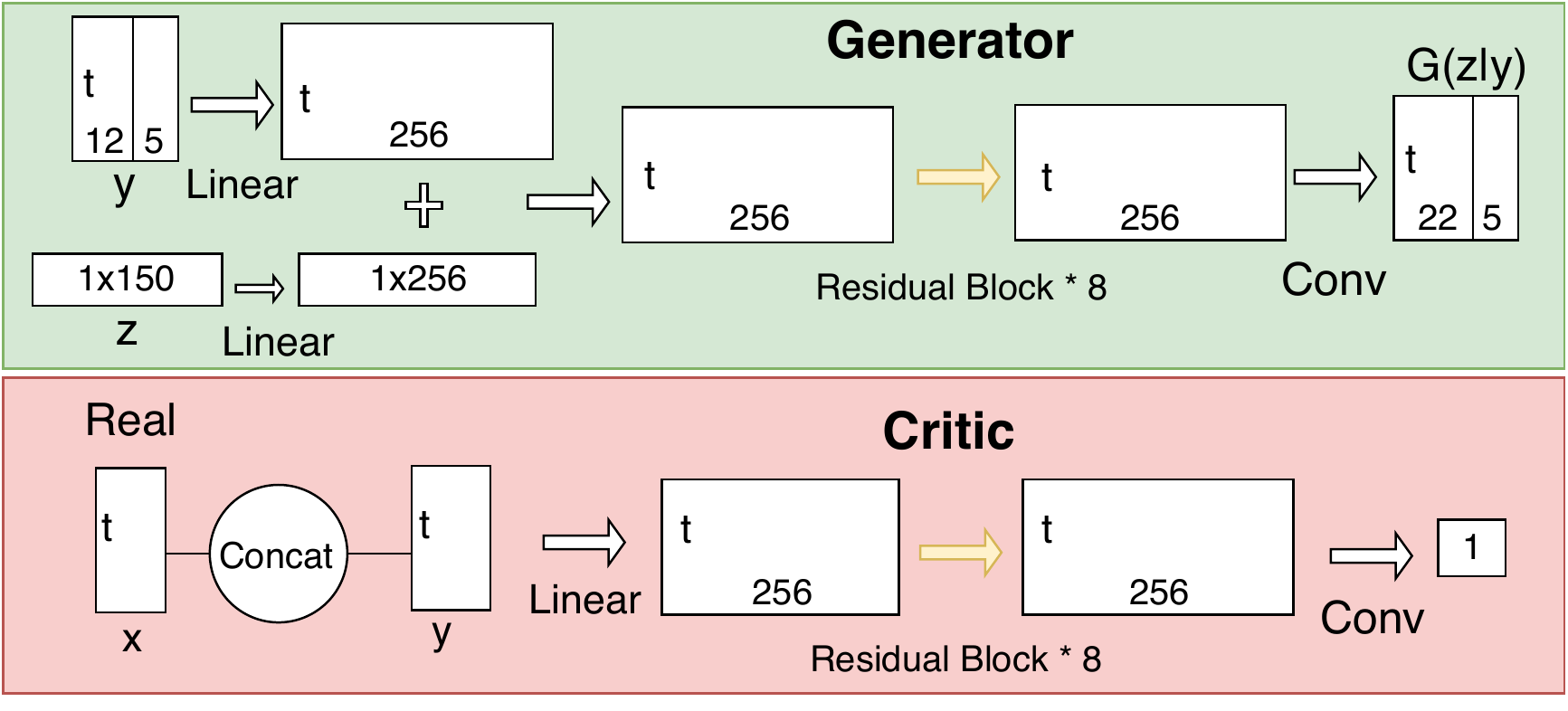}
\caption{Details of the generator and the critic network architectures.} 
\label{fig:network_GC}
\end{figure}

\begin{figure*}[t]
    \centering
    \begin{subfigure}{0.33\linewidth}
    \includegraphics[width=\linewidth]{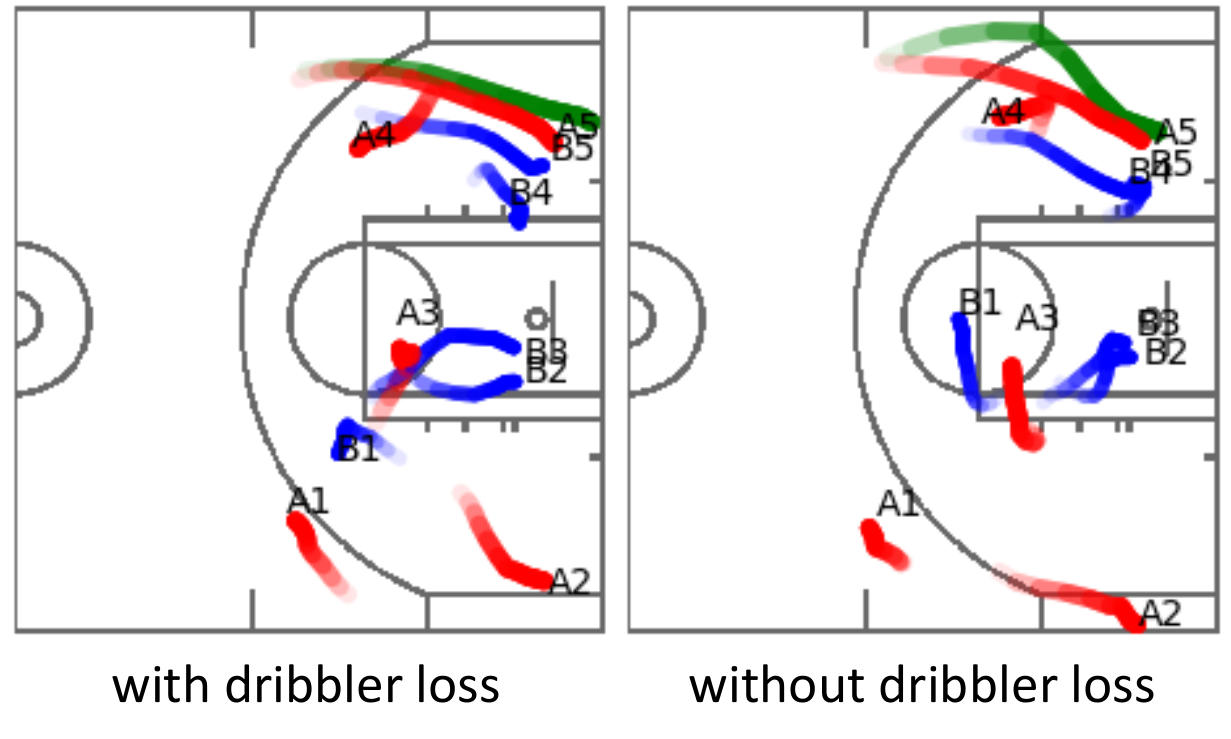}
    \caption{}
    \end{subfigure}
    \begin{subfigure}{0.33\linewidth}
    \includegraphics[width=\linewidth]{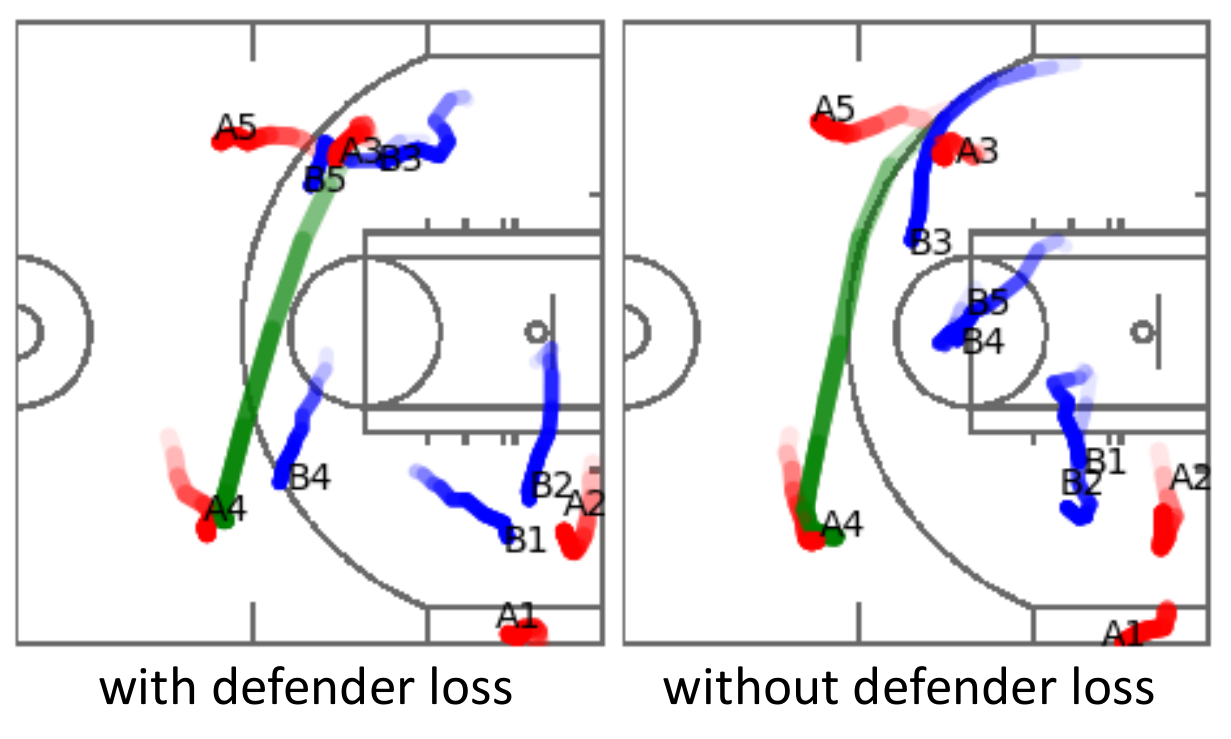}
    \caption{}
    \end{subfigure}
    \begin{subfigure}{0.33\linewidth}
    \includegraphics[width=\linewidth]{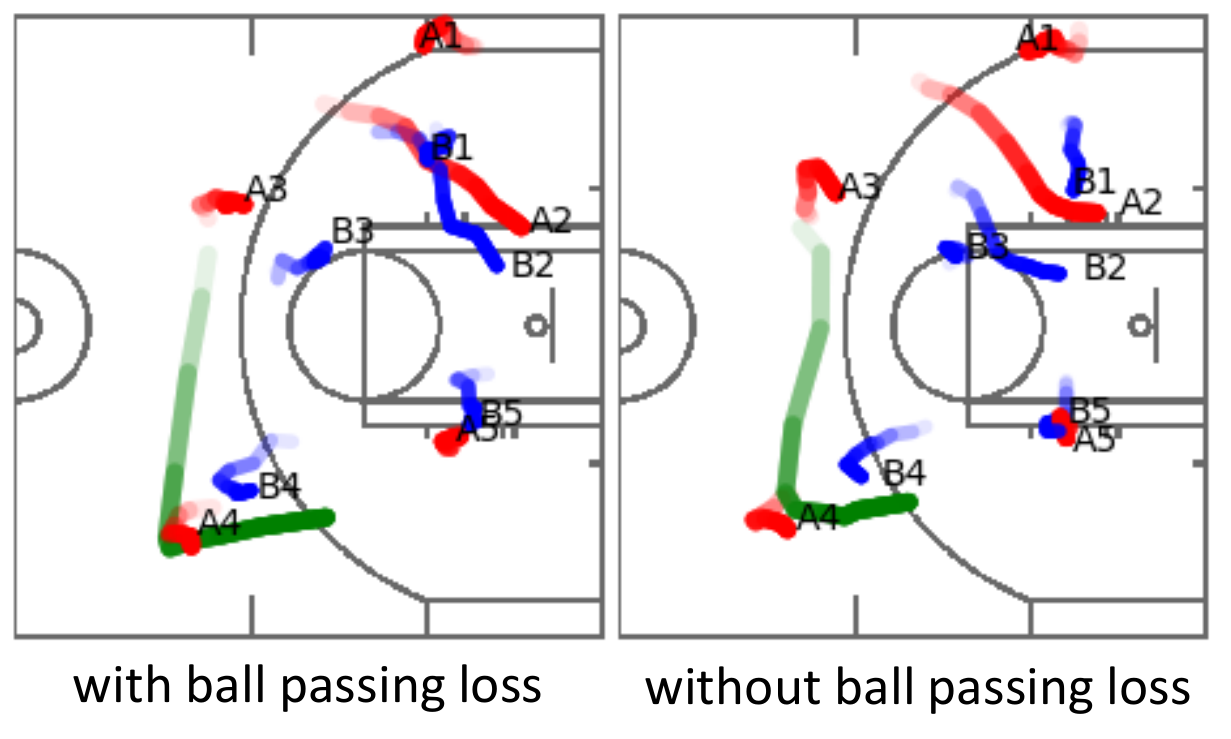}
    \caption{}
    \end{subfigure}
    \caption{The basketball plays are simulated by our network that is trained with and without the dribbler loss, straight line loss, and the wide open loss. Green, red, and blue trajectories show the movements of the ball, offensive players, and defensive players, respectively. (a) In the top right region, the ball deviates from the ball handler. (b) In the bottom left region, a wide open occurs when the ball is passed from A3 to A4. (c) In the middle left region, the ball passing trajectory is not straight.}
    \label{fig:loss_comp}
\end{figure*}

\textbf{Details of network architecture. } 
Figure \ref{fig:network_GC} shows the architectures of the presented generator $G$ and the critic network $C$. In the generator, it is noting that the condition $y$ and a latent noise $z$ are linearly projected to have the same size before the feature maps are combined through addition. This enables the latent noise to have global influence over the entire sequence, rather than individual frames. The combined feature map is passed into 8 residual blocks to construct the manifold feature. The output results from a down sampling convolution layer with stride 1 and a 5x5 kernel. In the critic, the real and fake pairs undergo the process similar to the generator. The network outputs a score to indicate the realism of a basketball play.

\subsection{Loss Functions}
We train the network by minimizing an adversarial loss, a dribbler loss, a defender loss, a ball passing loss, and an acceleration loss. The adversarial loss determines how real a generated basketball play is; the dribbler loss measures the distance between the ball and the dribbler; the defender loss penalizes the wide open; the ball passing loss ensures ball passing trajectories being straight lines; and finally, the acceleration loss prevents players from being too aggressive.

\textbf{Adversarial loss.}
We compute the adversarial loss by measuring the Wasserstein distance \cite{arjovsky2017wasserstein} between real and fake distributions. The output of the critic is a scalar score that determines how real a sample is. Formally, the loss is written as:
\begin{eqnarray}
\nonumber L_g(G, C) = & E_{y \sim P_{data}(Y), z \sim P_z(Z)} [C(G(z|y)|y)] \\
\nonumber  -& E_{x,y \sim P_{data}(X, Y)}[C(x|y)] \\
+&\lambda \times E_{\hat{x} \in P_{data} (\hat{X})} E[(\|\nabla_{\hat{x}} C(\hat{x}) \|_2 - 1)^2], 
\label{eq:general}
\end{eqnarray}
where $\hat{X}$ is the interpolation of the generated distribution and the real dataset, and $\lambda$ is the weight for the regularization term. We set $\lambda=10$ in the experiments.



\textbf{Dribbler loss. } 
The ball should not leave the dribbler when it is in possession. To prevent such a problem, we strive to minimize the Euclidean distance between the ball and the dribbler, where the dribbler is determined by the generated ball feature mentioned in Section \ref{sec:GAN_arch}. Let $f_i$ be the ball feature that indicates whether player $i$ possesses the ball, $\mathbf{p}_b$ and $\mathbf{p}_i$ be the ball position, and the position of player $i$, respectively, we minimize the loss
\begin{eqnarray}
L_{d} = \sum_t \sum_i f_i \times \big|\mathbf{p}_b - \mathbf{p}_i\big|
\label{eq:dribbler}
\end{eqnarray}
in each frame of the generated basketball play $G(z|y)$, where $t$ is the frame index, Note that $f_i = 1$ only if the player $i$ possesses the ball, and that the loss will be $0$ when the ball is passing or shooting. We show the results simulated by the network trained with and without this dribbler loss in Figure \ref{fig:loss_comp} (a).


\textbf{Defender loss. } 
A generated basketball play is unrealistic if the dribbler does not attempt to make scores when he/she is not defended. Hence, we minimize the Euclidean distance between the ball handler and the nearest defender that is between the player and the basket. To expect that the dribbler is defended as in the real basketball games, we minimize the following loss in each frame of the play:
\begin{eqnarray}
\nonumber L_{w} = \big|D(x) - D(G(z|y))\big|, \quad \textrm{where} \\
D(\bullet) = \sum_t (1 +\theta) \times \left(1 + |\mathbf{p}_n - \mathbf{p}_b|\right),
\label{eq:open}
\end{eqnarray}
$x$ is a random real basketball play, $\mathbf{p}_b$ is the ball position on the court, $\mathbf{p}_n$ is the closest defender to $\mathbf{p}_b$, $\theta$ is the angle determined by $\mathbf{p}_n-\mathbf{p}_b$ and a vector from the ball to the basket. We show the results simulated by the network trained with and without this defender loss in Figure \ref{fig:loss_comp} (b).

\textbf{Ball passing loss. }
Simulated ball passing trajectories in the XY-plane should be straight. To fulfill this requirement, we add a straight line loss to guide the network training. Let $\mathbf{p}_{b,i}$ be the ball position at frame $i$, and $\phi_i$ be the angle formed by the ball positions $\mathbf{p}_{b,i-1}$, $\mathbf{p}_{b,i}$, and $\mathbf{p}_{b,i+1}$. The loss is formulated as
\begin{eqnarray}
L_{b} = \sum_t \left(1 - \sum_i f_i \right) \times \phi_i.
\end{eqnarray}
Note that the loss is counted only when the ball is not in possession by players. We show the results simulated by the network trained with and without this ball passing loss in Figure \ref{fig:loss_comp} (c).


\textbf{Acceleration loss. }
Professional basketball players would prevent unnecessary movements on the court to save their physical energies. However, simulated players may behave aggressively to fulfill the above-mentioned requirements, such as avoiding wide open. Therefore, to achieve realism, we add an acceleration loss to keep simulated players moving as real players. Specifically, the loss is defined as
\begin{eqnarray}
L_{a} = \left|\mu(x) - \mu(G(z|y))\right|,
\end{eqnarray}
where $\mu(\bullet)$ is the mean acceleration of players in a set play.

By integrating the above mentioned losses, we minimize
\begin{eqnarray}
L = L_g + w \cdot (L_d + L_b + L_w + L_a)
\end{eqnarray}
to update the network hyper-parameters, where $w = |C(G(z|y))|$ is the critic score that helps balance the overall influence of the adversarial loss and our defined losses. 

\begin{figure*}[t]
    \centering
    \includegraphics[width=0.9\linewidth]{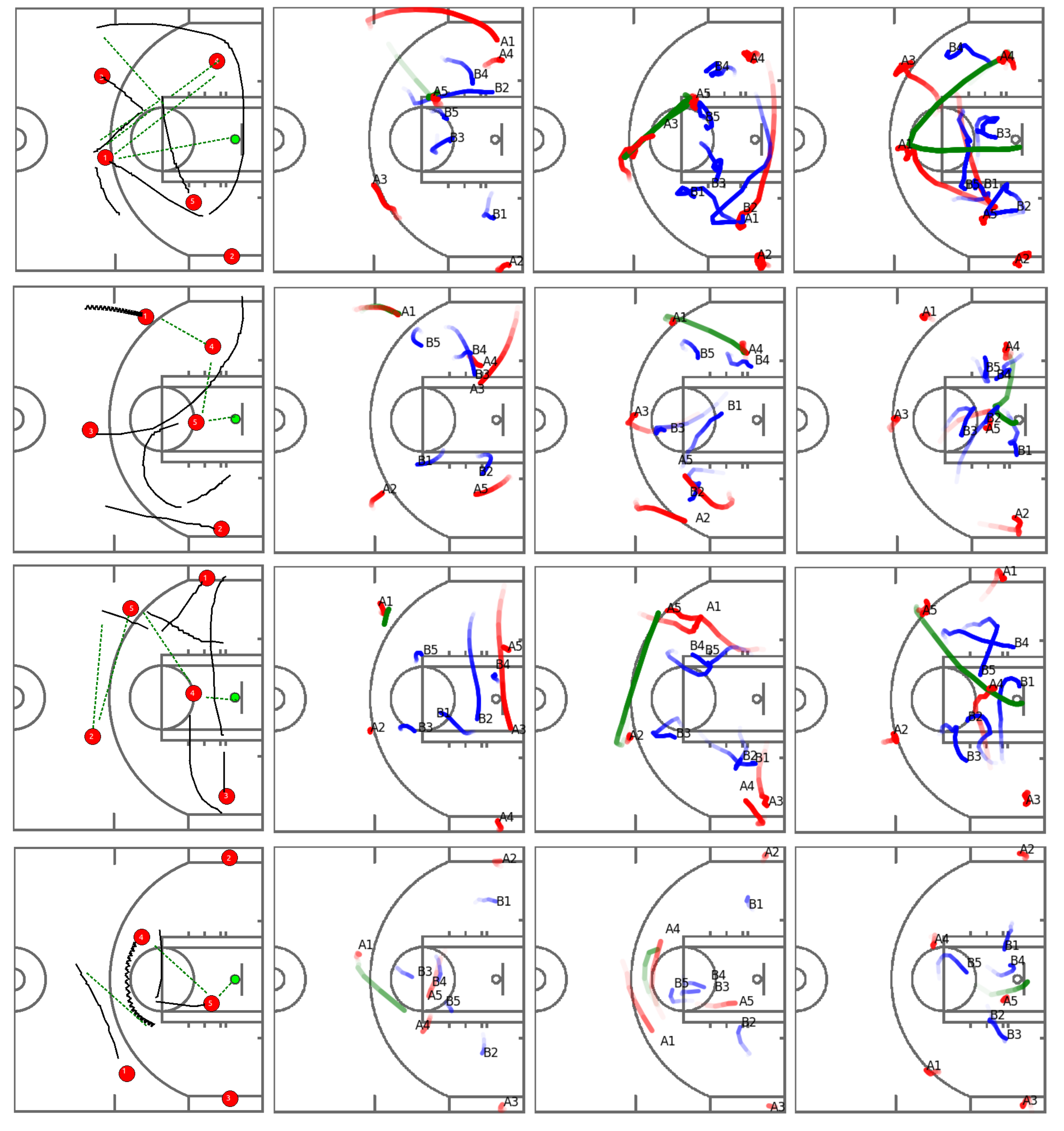}
\caption{We show basketball set plays simulated by our system based on four different offensive conditions. The tactics from top to bottom are snap down, hammer, flex, and elbow. The first column shows the sketches. The second to the fourth columns are the consecutive segments of our simulated results.}
\label{fig:results}
\end{figure*}

\subsection{Implementation Details}
We trained the network on a single NVIDIA GeForce 1080ti GPU. Adam optimizer \cite{kingma2014adam} with a learning rate 1e-4 and a batch size 128 with a fixed sequence length (t = 50) were used. In addition, the hyper-parameters of the network were initialized using Xavier \cite{glorot2010understanding}. Although the network is trained with a fixed sequence length, during testing phase variable length sequences can be fed into the network for basketball play simulation.

A three step strategy is implemented for obtaining the best results. First, the critic is pre-trained for 10 epochs with generator only trained for 1 iteration per epoch. This ensures that the critic is optimum to start with and feed meaningful gradients to the generator to improve upon. Then, we train the critic for 5 iterations before training the generator once, and with every 20 epochs we update the critic 10 iterations per generator iteration, as this strategy keeps the critic remaining strong. We stop the training process when the critics start overfitting.


\begin{table*}[t]
    \centering
    \begin{tabular}{|c|c|c|c|c|c|c|c|c|}
    \hline
        & \multicolumn{2}{c|}{Ball} & \multicolumn{2}{c|}{Offensive players} & \multicolumn{2}{c|}{Defensive players} & \multicolumn{2}{c|}{Ball-player relationship}   \\
    \cline{2-9}
        & velocity & acceleration & velocity & acceleration & velocity & acceleration & Ball-dribbler Dis & Ball-defender Dis  \\
    \hline
    Real & 9.52 (8.62) & 4.59 (5.47) & 5.01 (3.78) & 0.85 (0.94) & 4.03 (2.93) & 0.83 (0.92) & 1.55 (0.79) & 6.37 (2.75)        \\
    Ours & 7.43 (6.98) & 2.29 (3.38) & \textbf{4.93 (3.48)} & 1.17 (1.31) & \textbf{4.12 (2.87)} & 1.19 (1.36) & \textbf{1.53 (1.56)} & \textbf{6.17 (3.03)}        \\
    C-VAE & \textbf{7.65 (7.10)} & \textbf{2.60 (3.25)} & 4.68 (3.35) & \textbf{1.12 (1.37)} & 3.22 (2.16) & \textbf{0.97} (1.22) & 1.47 (1.43) & 6.09 (2.49) \\
    DEF  &  x  & x & x & x & 4.29 (3.50) & 1.45 (2.18) & x & 6.05 (2.29) \\
    \hline
    \end{tabular}
    \caption{Statistics of real and simulated basketball set plays. The 2$^{rd}$ to the 7$^{th}$ columns show the mean velocity and acceleration of the ball, offensive players, and defensive players, respectively. The 8$^{th}$ and 9$^{th}$ columns show the distances from the ball to the dribbler and to the closest defender, respectively. The number in each bracket indicates the standard deviation. We highlight the statistic of the simulated data closest to that of the real data to facilitate comparison.}
    \label{tab:statistic}
\end{table*}

\section{Results and Evaluations}
We tested the trained neural network on a variety of sketched offensive conditions, including those consisting of frequent ball passing and screening. After the ball and the offensive players' movement trajectories are sketched on the tactical board, our system will simulate the basketball set play. Figure \ref{fig:results} shows several examples simulated by using our system. The offensive and defensive players are termed as A1-A5, and B1-B5, respectively. Considering that visualizing basketball set plays by using still images is difficult, we recommend readers to watch our supplemental videos for a better understanding of the plays.

\textbf{Snap down. }
The set play involves getting the attention away from A1 as he cuts away from the play after passing the ball to A5. After the initial hand-off, a down screen is set for A1 by A4, where A1 gets free from his defender. Meanwhile, A5 and A3 pass back and forth once, before the ball is passed to A4 after the screen is set. Then, A5 proceeds to set screen for A1 to get an open shot, receiving a pass at the top of the key from A4. 

\textbf{Hammer. }
This example shows how to get A5 open. In the beginning, A5 set the Hammer screen (off the ball screen) for A2 on the weak side. Typically, A5 is a center and not known for hitting shots outside the restricted area. This leads to A5 defender (B5) to be prone to sag off and help either defending A2 when the screen is set, or protect the rim. This helps opens up for a jump shot or cut into the restricted area for an easy layup from a pass by A4.

\textbf{Flex. }
A flex cut occurs when a player in the corner receives a screen from another player that leads him into the restricted area for an easy layup or shot. In this example, two separate flex cuts occur, starting with A3 receiving a screen from A5. The first flex is a dummy, as A5 moves up to receive the ball. A3 then screens for A4 to flex cut into the restricted area for an easy layup. 

\textbf{Elbow. }
The offence is performed by two big men (center and power-forward) who utilize the pick-and-roll action to create an open. After A1 passes the ball to A4, A5 sets a screen for A4 on top of the key. Once the screen is set, A5 rolls to the basket, giving the options for A4 to shoot or pass the ball to A5 for an easy layup. As shown in this this example, an open is created after A4 receives 3 defenders attention. 

\subsection{Data Order}
Players' positions in a condition $y$ and the simulated basketball set play $G(z|y)$ are represented by a $t \times 10$ and a $t \times 20$ matrices, respectively. The order of these players strongly affects the convergence speed. Specifically, if the first player acts as a guard in a set play and as a forward in another, the network had to take more steps to learn the simulation of basketball set plays. To speed up the training process, we sort offensive players according to the mean distances between the players and the ball in a set play. Then, each defensive player was attached to the closest offensive player to obtain his order. This sorting strategy enables the players playing the same position to be likely arranged in the same column of the matrix. We realize that this sorting method was not optimal because players' movements on the court are complex. However, experimentally we observed that the overall training steps can be reduced by more than 2 times by the sorting.

\subsection{Model Comparison}
\begin{figure}[t]
    \centering
    \includegraphics[width=1\linewidth]{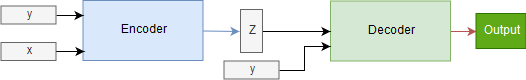}
    \caption{The architecture of C-VAE. During the training phase, both the condition $y$ and the real basketball set play $x$ are required. During the testing phase, the encoder can be neglected. Only the condition $y$ and a latent noise $z$ are fed into the decoder for reconstruction.} 
    \label{fig:CVAE}
\end{figure}

For quality evaluation purposes, two other generative models along with our system were trained and were compared with each other. Since conditional variational autoencoder (Figure \ref{fig:CVAE}) is known for good data reconstruction, we compared our system to this \emph{C-VAE} network. In addition, we compared our system to the work of Chen \textit{et al.} \cite{chen2018generating}, which takes real offensive set plays as input and outputs the corresponding defensive set plays. In the comparison, the input to this \emph{DEF} network was the same to the other networks. The results of the comparison are shown in our supplemental material.

\begin{figure}[t]
    \centering
    \includegraphics[width=1\linewidth]{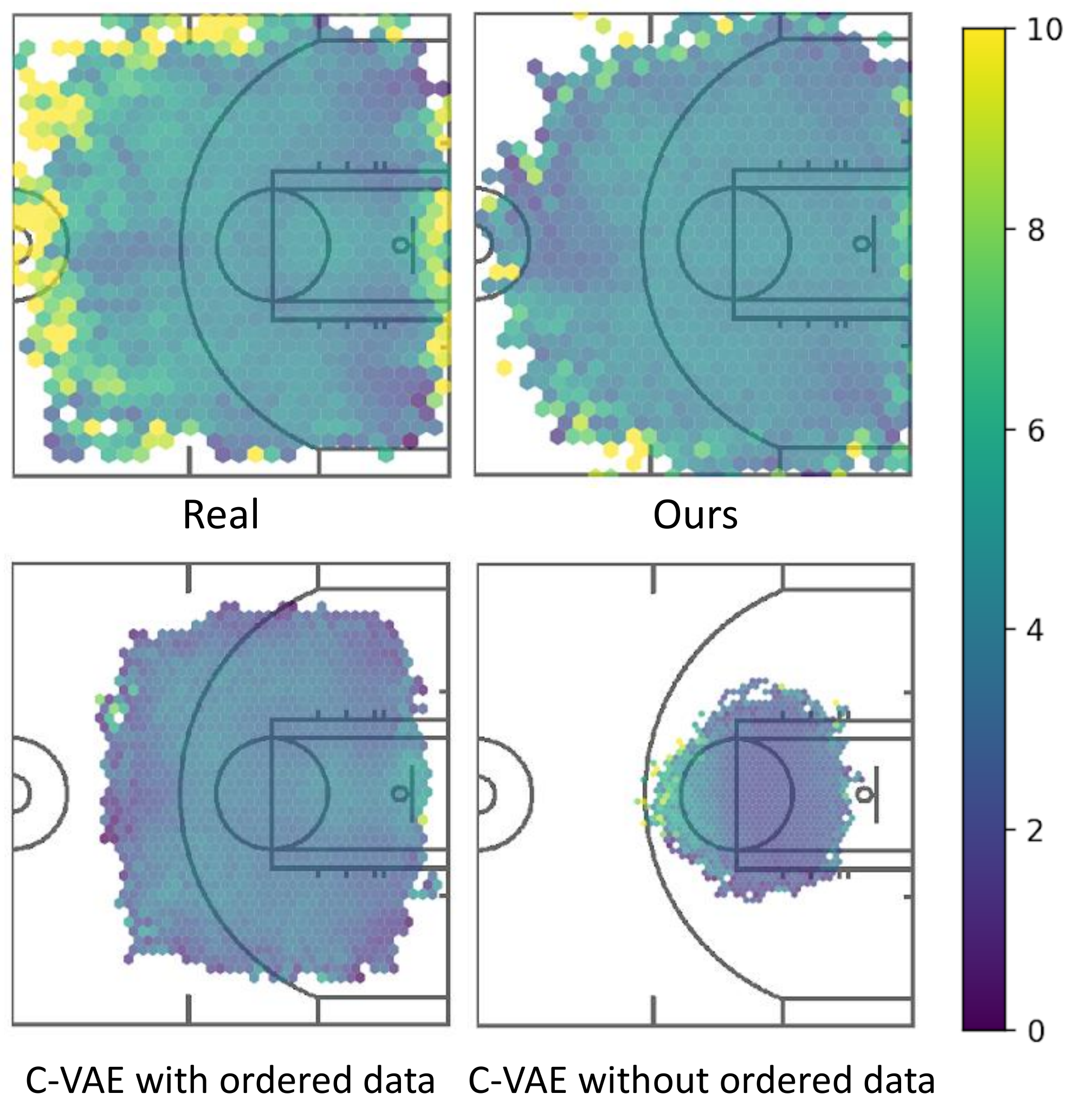}
    \caption{We show the distributions of defensive players on the court. The mean players' velocity at each position is represented by color. On the right is the transfer function.}
    \label{fig:velocity}
\end{figure}

\begin{table*}[t]
\centering
\begin{tabular}{|c|c|c|c|c|c|c|c|c|c|c|c|c|}
  \hline
    & \multicolumn{6}{c|}{Section 1: true or false} & \multicolumn{6}{c|}{Section 2: left or right} \\ \cline{2-13}
                  & Q1 & Q2 & Q3 & Q4 & Q5 &Q6 & Q1 & Q2 & Q3 & Q4 & Q5 &Q6  \\ \hline
    ORD           & 0.32 &	0.64 &	0.59 &	0.50 &	0.50 & 0.55 & 0.41 & 0.50 &	0.55 & 0.50 & 0.50 & 0.55 \\
    FAN           & 0.18 &	0.86 &	0.59 &	0.55 &	0.73 & 0.59 & 0.68 & 0.73 &	0.40 & 0.31 & 0.64 & 0.59 \\
    PRO           & 0.33 &	0.83 &	0.83 &	0.33 &	1.00 & 0.67 & 0.83 & 0.83 &	1.00 & 0.50 & 0.67 & 0.83 \\ \hline
\end{tabular}
\caption{We show the mean correct rates of each question answered by three different groups.} 
\label{tab:Userstudy}
\end{table*}

Table \ref{tab:statistic} shows the statistic of real and simulated basketball set plays that were experimented on the testing dataset. The ball, offensive players, and defensive players movements were evaluated. As indicated, our system performs better than C-VAE and DEF on players' movement velocity and the relationships between the ball and the related players. However, it makes the simulated players become aggressive as their accelerations are higher than the players simulated by other methods. We suspect the reason was that the discriminator determined the realism of a basketball set play by examining a sequence of player positions. The second derivatives of the player positions were considered less important. From the human perspective, this strategy is not harmful because the change of acceleration is less noticeable than the change of speed and position to humans. In addition, we observed that C-VAE could successfully reconstruct movements of the ball and the offensive players. The result was not surprising because of the hints from the given condition $y$. However, their simulated defensive players moved quite slowly on the court and tended to stay in two-point zone, as visualized in Figure \ref{fig:velocity}. The situation occurred because C-VAE is a supervised learning network. Finally, DEV was trained to simulate only defensive players. Besides the unrealistic offensive set plays, the simulated defensive set plays were also less realistic than ours.

It deserves nothing that the order of players' data strongly influenced the results of C-VAE. When the order was randomly assigned, the mean velocity of the defensive players dropped from 3.22 to 2.15. In addition, the players got closer to the restricted area (Figure \ref{fig:velocity}). By contrast, the GAN-based models did not suffer from the problem. In our experiments, although the training steps considerably increased under this circumstance, the quality of the simulations was held.

\subsection{User Study}
Conducting an objective measurement is a good way to evaluate the quality of the simulated basketball set plays. However, with subjective test results, one can be more convinced that our simulated basketball set plays resemble real natural movements. Therefore, we created a survey that consists of two sections, each with 6 questions. The simulated results were selected from the testing dataset. In section 1, participants were faced with 3 real and 3 fake basketball set plays and were asked to answer whether the play was real or fake. The real and fake plays were displayed in a random order to avoid bias. In section 2, both real and fake plays were compared side by side. Participants had to determine the play on the left or on the right was real. Similarly, the position of each play was randomly determined. All 50 participants that took the survey have experience in activities of watching and playing basketball. Participants were categorized into three groups, ORD, FAN and, PRO. ORD were participants who know basketball rules and seldom watch basketball games; FAN were participants who often watch NBA TV, and PRO were participants who joined clubs and actively played basketball. 

Table \ref{tab:Userstudy} shows the mean correct rate of each question. Overall, the mean correct rate to all of the binary tests were 56.17 \%. Among the three groups, PRO performed the best. 67\% and 78\% of the questions in Section 1 and 2 were correctly answered, respectively. Their answers to question 5 in section 1 and question 3 in section 2 were all correct. FAN was the second. 58\% and 56\% of the questions in Section 1 and 2 were correctly answered, respectively. ORD was the last. 51\% and 50\% of the questions in Section 1 and 2 were correctly answered, respectively. Their answers were close to random guess. In addition, the overall statistic was reasonable, as PRO was the group the most familiar with tactics and ORD was the least. We also observed that PRO could correctly answer more questions in Section 2 than in Section 1. While real and fake basketball plays were put together and compared side by side, they were likely to pick the right play.

\subsection{Limitations}
We present a network to simulate realistic basketball set plays according to the sketched offensive conditions. A noted limitation is that input sketches for training are simplified from real players' movement trajectories rather than being specified by users. Although experimentally we did not observe noticeable artifacts in the simulations based on user sketches, theoretically we cannot claim that the two types of sketches are the same. In addition, we observed that offensive players not given instructions will not leave their positioned region in the simulation. However, in reality the players react on their teammates actions. They may help create a wide open or receive the ball to make scores. On the other hand, defensive players were simulated without receiving any instructions. One idea to improve the system is conditioning defensive schemes by labels such as zone defence and man-to-man. As a result, players and coaches will be able to forecast different defensive reactions to their offense. Finally, the system imitates just general player movements when it synthesizes basketball set plays. In practice, it is beneficial for coaches to teach their players. In competition, however, the system can be improved further to imitate a higher-level of specific team or player movement behavior. By taking account of players' profiles, we could see different defensive approaches such as not pressuring a poor three-point shooter when he has the ball around the three-point line.

\section{Conclusions and future works}
We have presented a deep neural network to simulate basketball set plays according to sketched offensive conditions. The network is trained on NBA player tracking data and learns to construct details between sketches and real offensive players' movement trajectories, and the missing defensive set plays. To achieve realistic simulations, we introduced several loss functions to assist network training. Experiment results and comparisons demonstrated that simulated basketball set plays can be indistinguishable from real set plays. Accordingly, by using our system, players and coaches can rely on objective data analysis rather than subjective intuition, and foresee how the defensive team will react to their offensive tactic. The simulation of basketball set plays help players get insights from tactics and help coaches make crucial decisions. 

Our current network is trained on general basketball players. It can be fine-tuned to simulate basketball set plays that a specific team or player is likely to react. Is this case, the network will be particularly useful for coaches to design tactics and select players as the starting lineup against the opposing team. We thus plan to collect and prepare data for network fine tuning in the near future. We will also improve the usability of our system and release the codes/program for public use.




\section*{Acknowledgements}
We thank anonymous reviewers for their insightful comments and suggestions. We are also grateful to Prof. Chin-Jen Cheng for the valuable discussions, and all the participants who joined the user study. This work is partially supported by the Ministry of Science and Technology, Taiwan, under Grant No. 105-2221-E-009 -135 -MY3 and 107-2221-E-009 -131 -MY3.

\bibliographystyle{ACM-Reference-Format}
\bibliography{ref.bib}

\end{document}